\documentclass[twocolumn,showpacs,preprintnumbers,amsmath,amssymb]{revtex4}
\input psfig.sty

\usepackage{graphicx}
\usepackage{dcolumn}
\usepackage{bm}

\begin{document}

\title{Complementary Constraints on Brane Cosmology}

\author{J. S. Alcaniz{\footnote{alcaniz@on.br}}}

\affiliation{Departamento de Astronomia, Observat\'orio Nacional, 20921-400 Rio de
Janeiro - RJ, Brasil}

\author{Zong-Hong Zhu{\footnote{zhuzh@bnu.edu.cn}}}

\affiliation{Department of Astronomy, Beijing Normal University, Beijing 100875, China}

\date{\today}

\begin{abstract}

The acceleration of the expansion of the universe represents one of the major challenges to our current understanding of fundamental physics. In principle, to explain this phenomenon, at least two different routes may be followed: either adjusting the energy content of the Universe -- by introducing a negative-pressure dark energy -- or modifying gravity at very large scales -- by introducing new spatial dimensions, an idea also required by unification theories. In the cosmological context, the role of such extra dimensions as the source of the dark pressure responsable for the acceleration of our Universe is translated into the so-called brane world (BW)  cosmologies. Here we study complementary constraints on a particular class of BW scenarios in which the modification of gravity arises due to a gravitational \emph{leakage} into extra dimensions. To this end, we use the most recent Chandra measurements of the X-ray gas mass fraction in galaxy clusters, the WMAP determinations of the baryon density parameter, measurements of the Hubble parameter from the \emph{HST}, and the current supernova data. In agreement with other recent results, it is shown that these models provide a good description for these complementary data, although a closed scenario is always favored in the joint analysis. We emphasize that observational tests of BW scenarios constitute a natural verification of the role of possible extra dimensions in both fundamental physics and cosmology.
\end{abstract}

\maketitle

\section{Introduction}

The idea of a dark \emph{pressure}-dominated universe is usually associated
with unknown physical processes involving either new fields in
high energy physics or modifications of gravity at very large
scales. If one chooses to follow the former route, then the two
favorite candidates for this mysterious component are the energy
density stored on the true vacuum state of all existing fields in
the Universe, i.e., the cosmological constant ($\Lambda$),
and the potential energy density associated with a dynamical
scalar field ($\phi$), usually called dark energy or \emph{quintessence} (see,
e.g., \cite{revde} for some recent reviews on this topic). 

The second route in turn is predominantly related to the existence of extra
spatial dimensions, an idea that is required in various theories beyond the
standard model of particle physics, especially in theories for unifying gravity and
the other fundamental forces, such as superstring or M theories. In the
eleven-dimensional supergravity model of Ho\v{r}ava and Witten
\cite{witten}, for instance, the ordinary matter fields are assumed to be confined in a
submanifold
(brane) immersed in a higher dimensional space, usually named bulk. An important 
development of this idea was subsequently given by Randall and Sundrum \cite{rs} who
showed that, if
our three-dimensional world is embedded in a four-dimensional anti-de Sitter bulk, 
gravitational excitations are confined close to our submanifold, giving rise to the
familiar $1/{r^2}$ law of gravity (see also \cite{braneS} for further discussions).

From the observational viewpoint, the most important aspect asssociated with these
scenarios
resides on the fact that some of their versions can lead to a late time accelerating
universe, in agreement with supernovae observations \cite{rnew}. For this reason (and
also motivated by a possible explanation for the hierarchy problem from extra-dimension
physics), braneworld (BW)  cosmologies has
become a topic of much interest recently. Sahni and Shatnov \cite{ss}, for instance,
proposed a new class of BW models
which admit a
wider range of possibilities for the dark pressure than do the usual dark energy
scenarios. As shown in Ref. \cite{ss}, for a subclass of the parameter values the
acceleration of the Universe can be a transient phenomena, which could help reconcile
the supernova evidence for an accelerating universe with the requirements of
string/M-theory \cite{fis}. More recently, Maia {\it et al.} \cite{maia} showed that the
dynamics of a dark energy component parameterized by an equation of state $p = \omega
\rho$ (the so-called XCDM parametrization) can be fully described by the effect of the extrinsic curvature of a FRW universe
embedded into a five-dimensional, constant curvature de-Sitter bulk.

Another interesting scenario is the one proposed by Dvali {\it{et al.}}
\cite{dgp}, which we will refer to it as DGP model. It describes a self-accelerating
5-dimensional BW
model with a noncompact, infinite-volume extra dimension whose dynamics of
gravity is governed by a competition between a 4-dimensional Ricci scalar term, induced
on the brane, and an ordinary 5-dimensional Einstein-Hilbert action (see \cite{deff1}
for details). DGP models have been sucessfully tested
in many of their observational predictions, ranging from local gravity to cosmological
observations \cite{deff1,alc,alc1,alc2,lue} (see, however, \cite{avelino,Danswer}). From the theoretical viewpoint, the
consistency of their self-accelerating solutions is still a matter of debate in the
current literature (see, e.g., \cite{luty}).

This paper aims at placing new observational constraints on DGP models from the
current X-ray observations of rich clusters of galaxies and type Ia supernovae (SNe Ia)
data. We emphasize that this particular combination of observational data constitutes an
interesting and \emph{complementary} probe for testing the viability of cosmological
scenarios because while X-ray data are very effective to place limits on the clustered matter
(dark matter) the new SNe Ia sample tightly constrains the unclustered component (dark
pressure). To this end, we use the latest Chandra measurements of the X-ray gas mass
fraction in 26 galaxy clusters, as provided by
Allen {\it et al.} \cite{allen1} along with the most recent determinations of the
baryon density parameter, as given by the WMAP team \cite{wmap}, the latest
measurements of the Hubble parameter, 
provided by the \emph{HST} key
project \cite{freedman}, and the so-called \emph{gold} set of 157 SNe Ia, recently
published by Riess {\it et al.} \cite{rnew}. The analysis performed here,
therefore, updates the results of Ref. \cite{zhu} and complements the SNe Ia study
presented in Ref. \cite{nilza}.

\section{DGP models: basic expressions}

In DGP models, the presence of an infinite-volume extra dimension modifies the Friedmann 
equation in the following way \cite{deff1}
\begin{equation} 
\left[\sqrt{\frac{\rho}{3M_{pl}^{2}} + \frac{1}{4r_{c}^{2}}} +
\frac{1}{2r_{c}}\right]^{2} = H^{2}  
+ \frac{k}{R(t)^{2}},
\end{equation} 
where $\rho$ is the energy density of the cosmic fluid, $k = 0, \pm 1$
is the spatial curvature,   
$M_{pl}$ is the Planck mass and $r_c = M_{pl}^{2}/2M_{5}^{3}$ ($M_5$
is the 5-dimensional reduced  
Planck mass) is the
crossover scale defining the  
gravitational interaction among particles located on the brane, i.e., for scales $r <
r_c$,
the
gravitational force experienced by two punctual sources is the usual 4-dimensional
$1/r^{2}$ force whereas for distance scales $r > r_c$ the gravitational force
follows the 5-dimensional $1/r^{3}$ behavior. From the above equation we find that the
normalization
condition is now given by $\Omega_k + \left[\sqrt{\Omega_{\rm{r_c}}} +
\sqrt{\Omega_{\rm{r_c}} + \Omega_{\rm{m}}}\right]^{2} =  1$, 
where $\Omega_{\rm{m}}$ and $\Omega_k$ are, respectively, the matter
and curvature density  
parameters (defined in the usual way) and 
\begin{equation}
\Omega_{\rm{r_c}} = 1/4r_c^{2}H_o^{2},
\end{equation} 
is the density parameter associated with the crossover radius $r_c$. For
a flat universe, the  
normalization condition above reduces to $\Omega_{\rm{r_c}} = \left(1 - \Omega_{\rm{m}}
\right)^{2}/4$. As noticed in Ref. \cite{deff1}, the above described cosmology can be exactly reproduced by the standard one plus an additional dark energy component with a time-dependent  equation of state parameter $\omega^{eff}(z) = 1/{\cal{G}}(z, \Omega_{\rm{m}},\Omega_{\rm{r_c}}) - 1$, where ${\cal{G}}(z, \Omega_{\rm{m}},\Omega_{\rm{r_c}}) = \sqrt{4\Omega_{\rm{r_c}}/\Omega_{\rm{m}}x'^{-3} + 4})(\sqrt{\Omega_{\rm{r_c}}/\Omega_{\rm{m}}x'^{-3}} + \sqrt{\Omega_{\rm{r_c}}/\Omega_{\rm{m}}x'^{-3} + 1})$ and $x' = (1 + z)^{-1}$ (see \cite{wsps} for a discussion on time-dependent parametrizations for $\omega$).

To perform our statistical analysis
in the next section two observational quantities are of fundamental importance, namely, the angular diameter [$d_{\rm{A}}(z)$] and luminosity
distances [$d_{\rm{L}}(z)$] -- intrinsically related, in a homogeneous and isotropic
universe, by $d_{\rm{A}}(z)(1 + z)^{2} = d_{\rm{L}}(z)$. From the above equations, it
is straightforward to show that 
\begin{eqnarray}
d_{\rm{A}}^{\rm{DGP}}(z) & = & \frac{H_o^{-1}}{(1 +
z)|\Omega_k|^{1/2}} \times \\ \nonumber & & \times  
{S_{k}} \left[|\Omega_k|^{1/2} \int^{1}_{x'} {dx \over 
x^{2} {\cal{F}}(\Omega_{j}, x)}\right] ,
\end{eqnarray} 
where 
the function ${S_{k}}$ is defined by one of the following  
forms: $S_k(r) = \mbox{sinh}(r)$, $r$, and $\mbox{sin}(r)$, respectively, for open,
flat and closed geometries. The dimensionless function ${\cal{F}}(\Omega_{j}, x)$ is
given by 
\begin{equation}
{\cal{F}}(\Omega_{j},x) = \left[\Omega_k x^{-2} + \left(\sqrt{\Omega_{\rm{r_c}}} + 
\sqrt{\Omega_{\rm{r_c}} 
+ \Omega_{\rm{m}}x^{-3}}\right)^{2}\right]^{1/2},
\end{equation}
where $j$ stands for $m$, $r_c$ and $k$. As one may check, for $\Omega_k = 0$, the limit
$1/r_c \rightarrow 0$ ($\Omega_{\rm{r_c}} \rightarrow 0$) provides
\begin{eqnarray}
d_{\rm A}^{\rm{SCDM}}(z) = \frac{2H_o^{-1}}{(1 + z)^{3/2}}\left[(1 + z)^{1/2} -
1\right],
\end{eqnarray} 
which is the standard (SCDM) prediction for the angular diameter
distance $d_{\rm A}(z)$.

\section{Constraints on DGP models}

\subsection{$f_{\rm gas}$ versus redshift test}

The X-ray gas mass fraction test $[f_{\rm gas}(z)]$ was first introduced in Ref. \cite{sasa} and further
developed in Ref. \cite{allen} (see also \cite{ettori,lima,zhu,ratra}). This is based on the
assumption that rich clusters of galaxies are large enough to provide a fair
representation of the baryon and dark matter distributions in the Universe \cite{hogan}. Following
this assumption, the matter content of the Universe can be expressed as
the ratio between the baryonic content and the gas mass fraction, i.e., $\Omega_{\rm m}
\propto  \Omega_{\rm b}/f_{\rm gas}$. Moreover, as shown by Sasaki \cite{sasa}, since $f_{\rm gas} \propto
d_{\rm{A}}^{3/2}$ the model function can be defined as \cite{allen} 
\begin{eqnarray}
f_{\rm gas}^{\rm mod}(z) = \frac{ b\Omega_{\rm b}} {\left(1+0.19
\sqrt{h}\right) \Omega_{\rm m}} \left[ 2h \,
\frac{d_{\rm A}^{\rm{SCDM}}(z)}{d_{\rm A}^{\rm{DGP}}(z)} \right]^{1.5},
\end{eqnarray}
where the bias factor $b$ is a parameter motivated by gas dynamical simulations that
takes into account the fact that the baryon fraction in clusters is slightly depressed
with respect to the Universe as a whole \cite{simula}, the term $0.19\sqrt{h}$ stands
for the optically luminous galaxy mass in the cluster and the ratio ${d_{\rm
A}^{\rm{SCDM}}(z_{\rm i})}/{d_{\rm A}^{\rm{DGP}}(z_{\rm i})}$ accounts for deviations
in the geometry of the universe (here modelled by the DGP model) from the default
cosmology used in the observations, i.e., the SCDM model (see \cite{allen1,allen} for
more observational details). 

\begin{figure}
\centerline{\psfig{figure=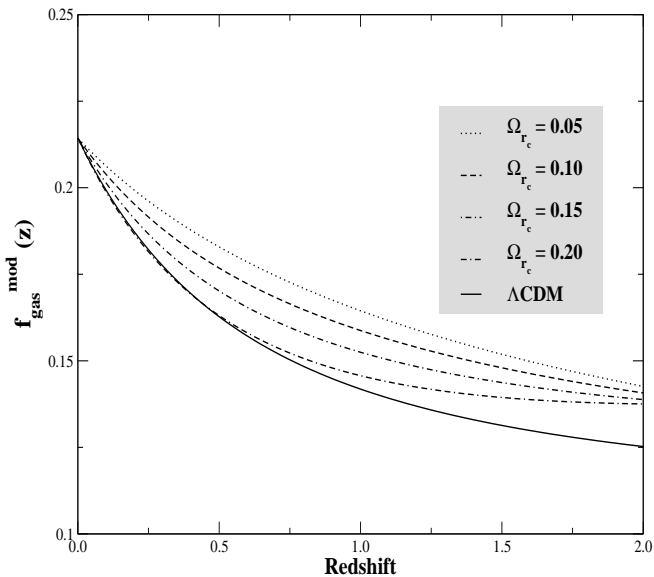,width=3.5truein,height=3.2truein,angle=-90}
\hskip 0.1in}
\caption{The model function $f_{\rm gas}^{\rm mod}$ as a function of the redshift for some selected values of $\Omega_{\rm{r_c}}$ and $\Omega_{b}h^{2} = 0.0224$, $b = 0.824$ and $h = 0.72$. The value of $\Omega_m$ is fixed at 0.3, as suggested by dynamical estimates \cite{calb}. The current concordance model, i.e., a flat scenario with $70\%$ of the critical energy density dominated by a cosmological constant, is also shown for comparison.}
\end{figure}

In order to constrain the cosmological parameters $\Omega_m$ and $\Omega_{\rm{r_c}}$ we
use the latest Chandra measurements of the X-ray gas mass fraction in 26 dynamically
relaxed galaxy clusters ($0.07 < z < 0.9$), as provided by Allen {\it et al.}
\cite{allen1}. The confidence intervals are determined by using the  conventional
$\chi^{2}$ minimization with the Gaussian priors on the baryon density parameter,
$\Omega_bh^{2} = 0.0224 \pm 0.0009$ \cite{wmap}, on the Hubble parameter, $h = 0.72 \pm 0.08$ \cite{freedman}, and on the bias factor, $b = 0.824 \pm 0.089$
\cite{simula}, i.e.,
\begin{eqnarray}
\chi^2_{f_{\rm gas}} & = &\sum_{i = 1}^{26}
\frac{\left[f_{\rm gas}^{\rm mod}(z_{\rm i})- f_{\rm gas,\,i}
\right]^2}{\sigma_{f_{\rm gas,\,i}}^2} 
+ \left[\frac{\Omega_{b}h^{2} - 0.0224}{0.0009}\right]^{2} 
+ \nonumber \\ & & 
\quad  \quad \quad + \left[\frac{h - 0.72}{0.08}\right]^{2} + \left[\frac{b -
0.824}{0.089}\right]^{2}.
\end{eqnarray}
In the above expression, $f_{\rm gas}^{\rm mod}(z_{\rm i})$ is given by Eq. (6) and $f_{\rm gas,\,i}$ is
the observed values of the X-ray gas mass fraction with errors $\sigma_{f_{\rm
gas,\,i}}$. To plot two-dimensional confidence contours, we have projected our
five-dimensional parameter space (defined by the vector $\vec{p} = \{\Omega_m, \Omega_{\rm{r_c}}, h,
\Omega_bh^{2}, b\}$) into the plane $\Omega_m - \Omega_{\rm{r_c}}$, which is similar to marginalize over the parameters $h$, $\Omega_bh^{2}$ and $b$ by defining the probability distribution function ${\cal{L}}(\Omega_{m}, \Omega_{\rm{r_c}}) = \int{dh d(\Omega_bh^{2}) db e^{-\chi^{2}/2}}$ (see, e.g., \cite{ratra}).

    Figure 1 shows the behavior of $f_{\rm gas}^{\rm mod}$ as a function of the redshift for some selected values of $\Omega_{\rm{r_c}}$ and $\Omega_{b}h^{2} = 0.0224$, $b = 0.824$ and $h = 0.72$. The value of $\Omega_m$ is fixed at 0.3, as suggested by dynamical estimates on scales up to about $2h^{-1}$ Mpc \cite{calb}. For the sake of comparison, the current favored cosmological model, namely, a flat scenario with $70\%$ of the critical energy density dominated by a cosmological constant ($\Lambda$CDM), is also shown. We note from this figure that with a larger sample of high-$z$ X-ray measurements ($z \geq 1$) it will be possible to completely distinguish between DGP and $\Lambda$CDM scenarios. In Fig. 2a we present the first results of our statistical analysis. There, it is shown the confidence regions ($68.3\%$, $95.4\%$ and $99.7\%$) in the plane $\Omega_m - \Omega_{\rm{r_c}}$ by considering the 26 X-ray data points described above. The best-fit parameters for this analysis are $\Omega_m = 0.29$ and $\Omega_{\rm{r_c}} = 0.31$ with $\chi^2/\nu \simeq 1.07$ ($\nu$ is defined as degrees of freedom). In particular, this relative value of $\chi^2$ is slightly larger than the one found in Ref. \cite{allen1} for the $\Lambda$CDM case with arbitrary curvature, i.e., $\chi^2/\nu \simeq 1.02$. At $95.4\%$ confidence limit (c.l.) we obtain the intervals $0.23 \lesssim \Omega_m \lesssim 0.37$ and $0.25 \lesssim \Omega_{\rm{r_c}} \lesssim 0.37$ whereas by restricting our analysis to the flat case, we note that the
data favour a lower value of the matter density parameter, i.e., $\Omega_{\rm{m}} \simeq
0.23$ ($\Omega_{\rm{r_c}} \simeq 0.148$) with  $\chi_{min}^2/\nu \simeq
1.2$. Such a value, however, is inside the 1$\sigma$ interval of the WMAP and other recent 
estimates of the quantity of matter in the Universe,
$\Omega_{\rm{m}} = 0.27 \pm 0.04$ \cite{wmap}, and therefore does not imply any possible
conflit between the predictions of the model and the independent measurements of
$\Omega_{\rm{m}}$ (see \cite{avelino,Danswer} for a discussion on this point). 

\begin{figure*}
\centerline{\psfig{figure=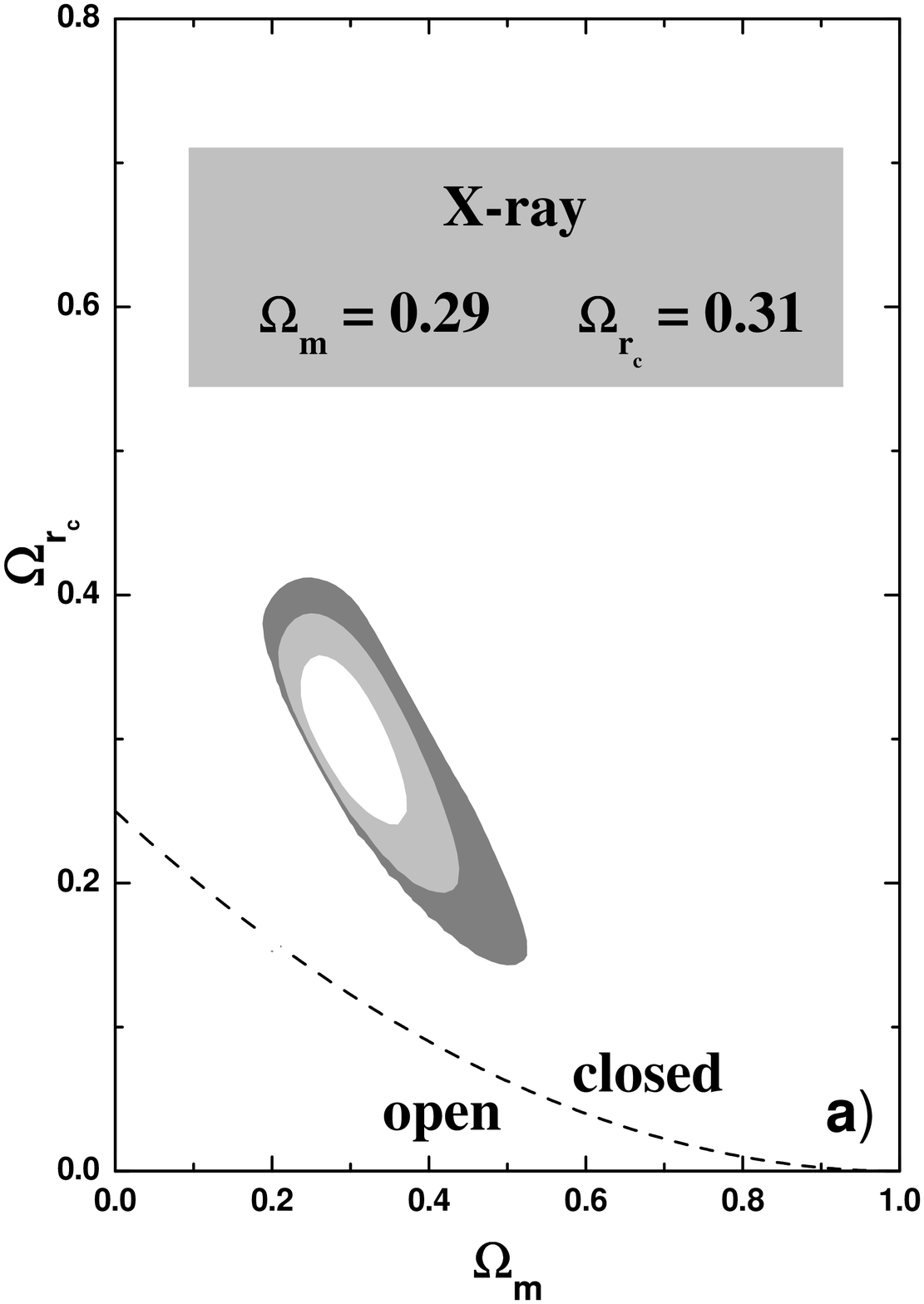,width=2.5truein,height=3.2truein}
\psfig{figure=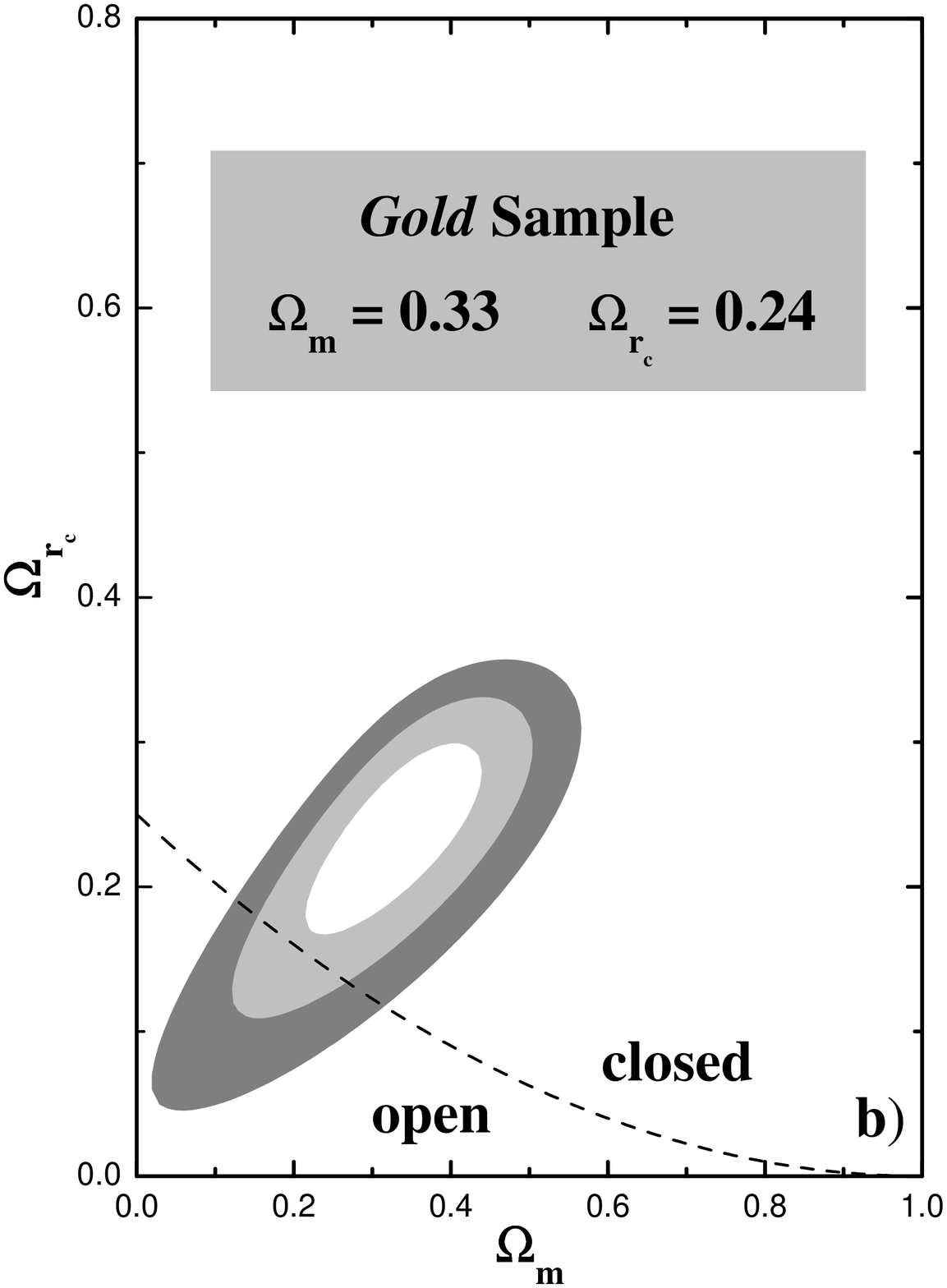,width=2.5truein,height=3.2truein}
\psfig{figure=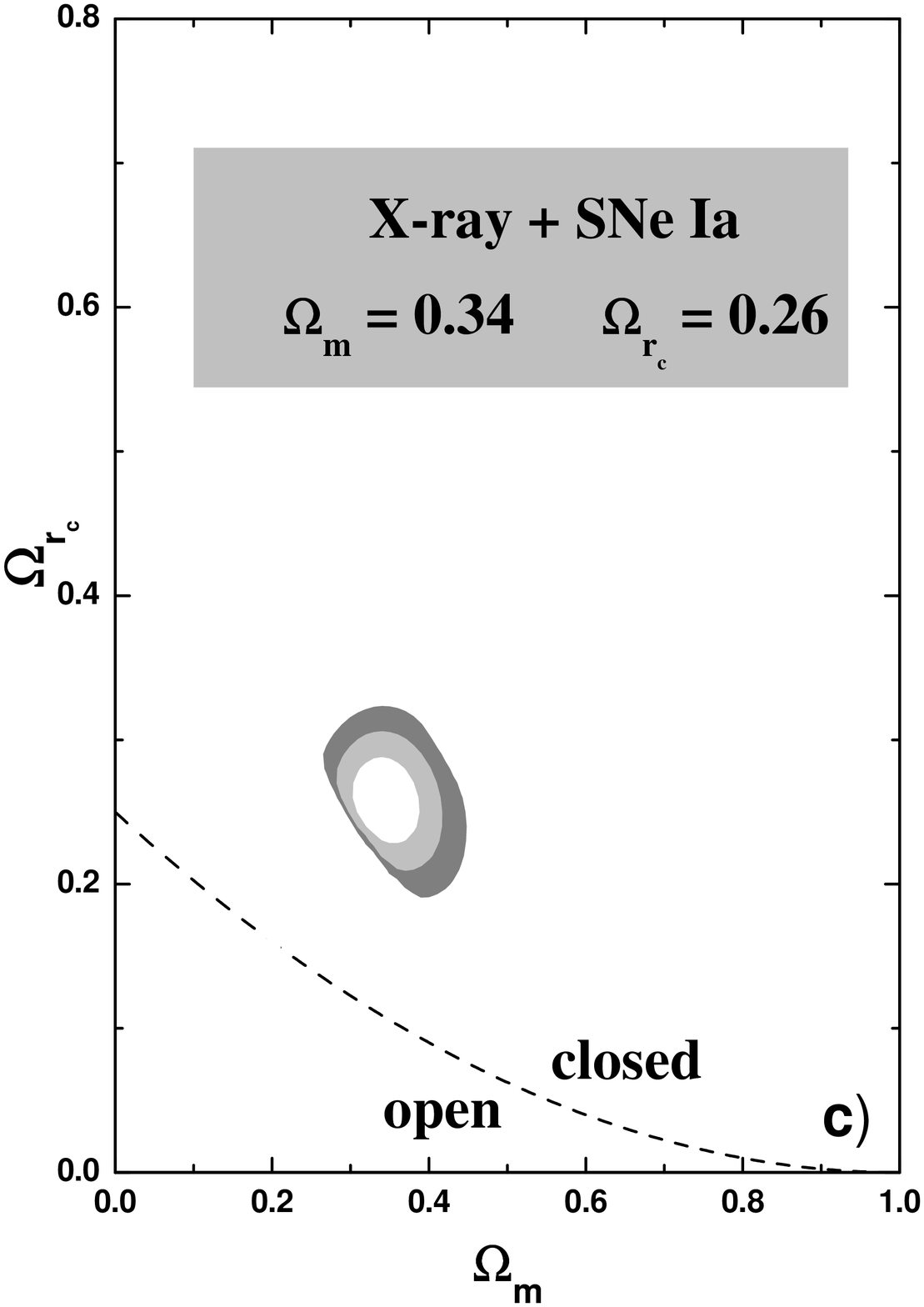,width=2.5truein,height=3.2truein}
\hskip 0.1in}
\caption{The results of our statistical analyses: Confidence regions ($68.3\%$, $95.4\%$ and $99.7\%$) in the $\Omega_{\rm{m}} - \Omega_{\rm{r_c}}$ plane by considering: {\bf{a)}} the latest Chandra measurements of the X-ray gas mass fraction in 26 galaxy clusters ($0.07 < z < 0.9$) plus determinations of the baryon density parameter and measurements of the Hubble parameter; {\bf{b)}} the so-called \emph{Gold} sample of Riess et al. \cite{rnew} -- taken from \cite{nilza} and shown here for the sake of comparison. Note that X-ray and SNe Ia data provide orthogonal statistics in the plane $\Omega_m - \Omega_{\rm{r_c}}$; {\bf{c)}} Joint X-ray + SNe Ia + $\Omega_{b}h^{2}$ + $H_o$ analysis. This combination of data provides $\Omega_m = 0.34^{+0.07}_{-0.05}$ and $\Omega_{\rm{r_c}} = 0.26 \pm 0.04$ (at $95.4\%$ c.l.) and clearly favors a closed universe.}
\end{figure*}

\subsection{Complementary Constraints}

By comparing our Fig. 2a with Fig. 2 of Ref. \cite{nilza} (also shown here as Fig. 2b for the sake of comparison), it is clear that X-ray and SNe Ia data provide orthogonal statistics in the plane $\Omega_m - \Omega_{\rm{r_c}}$. This, therefore, suggests that possible degenenacies between these parameters may be broken by combining these two data sets in a joint statistical analysis. To perform such an analysis we make use of the so-called \emph{gold} SNe Ia sample of Riess et al. \cite{rnew}, which consists of 157 events distributed over the redshift interval $0.01
\lesssim z \lesssim 1.7$. This particular sample constitutes the selection of the best observations made so far by the two supernova search teams along with 16 new events observed by the \emph{HST}. Our approach for SNe Ia data is based on Ref. \cite{nilza} (see also \cite{sne}).

\begin{table}
\caption{Recent estimates of the crossover radius $r_c$}
\begin{ruledtabular}
\begin{tabular}{lcr}
Method&Reference&$r_{c}$\footnote{In units of $H_o^{-1}.$}\\
\hline \hline \\
SNe Ia + CMB & \cite{deff1} & $1.4$\\
SNe Ia & \cite{avelino} & $1.4$\\
Angular size & \cite{alc} & $0.94$\\
Gravitational lenses & \cite{alc1} & $1.76$\\
High-$z$ galaxies & \cite{alc2} & $\leq 2.04$\\ 
SNe Ia + X-ray & \cite{zhu} & $1.09$\\
SNe Ia\footnote{Most recent SNe Ia data} + $\Omega_m$& \cite{nilza} & $1.09$\\
X-ray\footnote{Most recent X-ray data}: & & \\
\quad     Arbitrary curvature & This paper &
 $0.90$\\
\quad     Flat case & This paper & $1.30$\\
SNe Ia$^{b}$ + X-ray$^{c}$: & & \\
\quad     Arbitrary curvature & This paper & $0.98$\\
\quad     Flat case & This paper & $1.28$\\
\end{tabular}
\end{ruledtabular}
\end{table}

The results of our joint analysis are shown in Figure 2c. Note that the combination of these data sets leads to tight constraints on the $\Omega_m - \Omega_{\rm{r_c}}$ plane. A comparison with Fig. 3 of Ref. \cite{zhu} (which used the then available 172 SNe Ia taken from Tonry et al. \cite{tonry} plus 9 X-ray clusters from Allen et al. \cite{allen}) shows that the current analysis reduces considerably the area corresponding to the confidence intervals. As expected, the parameter space now is reduced relative to our former analysis, with the best-fit scenario occurring at $\Omega_m = 0.34$ and $\Omega_{\rm{r_c}} = 0.26$ ($\chi^2/\nu = 1.11$). These values  correspond to an accelerating universe with deceleration parameter $q_o = -0.8$ and a total expanding age of $t_o \simeq 10.1h^{-1}$ Gyr. At $95.4\%$ c.l. we found the following intervals: $0.29 \lesssim \Omega_m \lesssim 0.41$ and $0.22 \lesssim \Omega_{\rm{r_c}} \lesssim 0.30$. Note that this 2$\sigma$ interval for $\Omega_{\rm{r_c}}$ leads to an estimate of the crossover scale $r_c$ in terms of the present Hubble radius $H_o^{-1}$ [see Eq. (2)], i.e.,
\begin{equation}
r_c  = 0.98^{+0.08}_{-0.07}H_o^{-1},
\end{equation}
which is slightly smaller than (but in agreement with) the value obtained in Ref. \cite{zhu}, $r_c  = 1.09^{+0.29}_{-0.16}H_o^{-1}$ (99$\%$ c.l.). If we restrict our analysis to a flat geometry, i.e., by imposing the normalization condition $\Omega_{\rm{r_c}} = (1 - \Omega_{\rm{m}})^{2}/4$, we obtain $\Omega_m \simeq 0.22$ ($\Omega_{\rm{r_c}} \simeq 0.152$), which corresponds to a $9.6h^{-1}$-Gyr-old universe with $q_o = -0.45$ and $r_c \simeq 1.28H_o^{-1}$.

At this point we compare our estimates of the crossover scale $r_c$ with other recent determinations of this quantity from independent methods. We note a reasonable agreement among them. For instance, in Ref. \cite{deff1} Deffayet et al. used the then available SNe Ia + CMB data to find $r_c \simeq 1.4H_o^{-1}$ for a flat model with $\Omega_m = 0.3$, in agreement with Avelino and Martins \cite{avelino} who used a larger sample of 92 SNe Ia events. The current measurements of the angular size of high-$z$ sources require an accelerating universe with a crossover radius of the order of $r_c \simeq 0.94H_o^{-1}$ \cite{alc} whereas the statistics of gravitationally lensed quasars implies $r_c \simeq 1.76H_o^{-1}$ \cite{alc1}. The less concordant (but not in desagreement) among the current estimates for $r_c$ comes from age measurements of high-$z$ galaxies, which requires $r_c \lesssim 2.04H_o^{-1}$ \cite{alc2}. These results, along with the main estimates of the present paper, are summarized in Table I.

\section{Final Remarks}
                                                                                
There is increasing evidence for an accelerating universe from various astronomical observations. However, understanding the acceleration mechanism based on fundamental particle physics is still one of the most important challenges in modern cosmology. An unknown dark energy component with negative pressure has usually been invoked  as the most feasible mechanism for the acceleration although effects arising from extra dimension physics can mimic dark energy through a modification on the Friedmann equation \cite{braneS,ss,maia,card}. In this paper, we have focused our attention on a specific self-accelerating five-dimensional braneworld scenario, the so-called DGP model \cite{dgp}.
                                                                                
We analyzed the scenario by using the most recent Chandra measurements of the X-ray gas mass fraction in galaxy clusters, the WMAP determinations of the baryon density parameter, measurements of the Hubble parameter from the HST, and the current supernova data. As shown, the model provides a good description for the so-called gold SNeIa  sample of Ref. \cite{rnew}. However, in order to explain the X-ray data of clusters in the framework of DGP model, a closed universe is necessary (a similar result is also obtained by fitting SNeIa data to the standard $\Lambda$CDM model. In this case, CMB data is generally used to match a flat universe.).
The combination of the above mentioned data sets leads to very tight constraints on the $\Omega_m - \Omega_{\rm{r_c}}$ plane. At $95.4\%$ c.l. we found the following intervals: $0.29 \lesssim \Omega_m \lesssim 0.41$ and $0.22 \lesssim \Omega_{\rm{r_c}} \lesssim 0.30$, which gives a closed universe with the curvature of $-0.39 \lesssim \Omega_k \lesssim -0.18$. It is worth mentioning that we might make heavy use of the X-ray gas mass fraction in clusters, which further prefers to a closed universe in DGP model. This kind of analysis depends on the assumption that the $f_{\rm gas}$ values should be invariant with redshift, which has been criticised by a minority of workers in the field. For example, a recent comparison of distant clusters observed by XMM-Newton and Chandra satellites with available local cluster samples seems to indicate a possible evolution of the $M$--$T$ relation with redshift, which may be indicating that the  standard paradigm on cluster gas physics needs to be revised \cite{clusters}. Therefore, to pin down the DGP model from this kind of observations, a more detailed study on X-ray gas mass fraction in galaxy clusters is necessary. 

{\bf Acknowledgments:} The authors are very grateful to S. W. Allen for sending his compliation of the X-ray data and to G. S. Fran\c{c}a for valuable discussions on the manuscript. JSA is supported by CNPq (62.0053/01-1-PADCT III/Milenio). Z.-H. Zhu acknowledges support from the National Natural Science Foundation of China and the National Major Basic Research Project of China (G2000077602).


\begin{thebibliography}{30}

\bibitem{revde}  V. Sahni and A. Starobinsky, Int. J. Mod. Phys. {\bf{D9}}, 373 (2000); J. E. Peebles and B. Ratra Rev. Mod. Phys. {\bf{75}}, 559 (2003); T.
Padmanabhan, Phys. Rept. {\bf{380}}, 235 (2003); J. A. S. Lima, Braz. J. Phys. {\bf{34}}, 194  (2004). astro-ph/0402109

\bibitem{witten} P. Ho\v{r}ava and E. Witten, Nucl. Phys. {\bf{B460}}, 606 (1996); Nucl. Phys. {\bf{B475}}, 94 (1996).

\bibitem{rs}  L. Randall and R. Sundrum, Phys. Rev. Lett. {\bf{83}}, 3370 (1999); Phys. Rev. Lett. {\bf{83}}, 4690 (1999)

\bibitem{braneS}  P. Bin\'etruy, C. Deffayet, and D. Langlois, Nucl. Phys. {\bf{B565}}, 269 (2000); J. M. Cline, C. Grojean, and G. Servant, Phys. Rev. Lett. {\bf{83}}, 4245 (1999); T. Shi- romizu, K. Maeda, and M. Sasaki, Phys. Rev. {\bf{D62}}, 024012 (2001); V. Sahni, M. Sami, and T. Souradeep, Phys. Rev. {\bf{D65}} 023518 (2002); L. Randall, Science {\bf{296}}, 1422 (2002); H. J. Mosquera Cuesta, A. Penna-Firme, and A. P\'erez-Lorenzana, Phys. Rev. {\bf{D67}}, 087702 (2003)


\bibitem{rnew} A. G. Riess et al., \apj {\bf{607}} 665 (2004)

\bibitem{ss} V. Sahni and Y. Shtanov, IJMP {\bf{D11}}, 1515 (2002); JCAP {\bf{0311}},
014 (2003)

\bibitem{fis} W. Fischler et al., JHEP {\bf{0107}}, 003 (2001)

\bibitem{maia} M. D. Maia et al., astro-ph/0403072 (2004)

\bibitem{dgp} G. Dvali, G. Gabadadze and M. Porrati, Phys. Lett. {\bf{B485}}, 208 (2000)

\bibitem{deff1} C. Deffayet et al., Phys. Rev. {\bf{D66}}, 024019 (2002)

\bibitem{alc} J. S. Alcaniz, Phys. Rev. D {\bf{65}}, 123514 (2002). astro-ph/0202492

\bibitem{alc1} D. Jain, A. Dev and J. S. Alcaniz, Phys. Rev. {\bf{D66}}, 083511 (2002).
astro-ph/0206224

\bibitem{alc2} J. S. Alcaniz, D. Jain and A. Dev, Phys. Rev. {\bf{D66}}, 067301 (2002).
astro-ph/0206448

\bibitem{lue} C. Deffayet, G.  Dvali and G. Gabadadze, Phys. Rev. {\bf{D65}}, 044023
(2002);  A. Lue, Phys. Rev. {\bf{D67}}, 064004 (2003);  A. Lue and G. D. Starkman, Phys. Rev. {\bf{D67}}, 064002 (2003);  A. Lue, R. Scoccimarro and G. D. Starkman,  astro-ph/0401515

\bibitem{avelino} P. P. Avelino and C. J. A. P. Martins, ApJ {\bf{565}}, 661 (2002)

\bibitem{Danswer} C. Deffayet, G. R. Dvali and G. Gabadadze, astro-ph/0106449

\bibitem{luty} M. A. Luty, M. Porrati and R. Rattazzi, JHEP {\bf{0309}} 029 (2003); A.
Nicolis and R. Rattazzi, hep-th/0404159.

\bibitem{allen1} S. W. Allen et al., MNRAS (in press). astro-ph/0405340 (2004)


\bibitem{wmap}  D. N. Spergel et al., \apj Suppl. {\bf{148}}, 175 (2003)

\bibitem{freedman} W. L. Freedman {et al.}, \apj {\bf{553}}, 47 (2001)

\bibitem{zhu} Z.-H. Zhu and J. S. Alcaniz, \apj (in press). astro-ph/0404201

\bibitem{nilza} J. S. Alcaniz and N. Pires, Phys. Rev. {\bf{D70}}, 047303 (2004). astro-ph/0404146

\bibitem{wsps} Y. Wang and P. M. Garnavich, \apj {\bf{552}} 445 (2001); P.S. Corasaniti {\it et al.}, astro-ph/0406608; Y. Wang and M. Tegmark, Phys. Rev. Lett. {\bf{92}}, 241302-1 (2004); H. K. Jassal, J. S. Bagla, and T. Padmanabhan, astro-ph/0404378; D. Jain, J. S. Alcaniz and A. Dev., astro-ph/0409431; B. Bassett, P. S. Corasaniti and M. Kunz, astro-ph/0407364

\bibitem{sasa} S. Sasaki, PASJ, {\bf{48}}, L119 (1996)

\bibitem{allen} S. W. Allen, S. Ettori, A. C. Fabian, MNRAS, {\bf{324}}, 877 (2002); S. W. Allen, R. W. Schmidt, A. C. Fabian, MNRAS, {\bf{334}}, L11 (2002).

\bibitem{ettori} S. Ettori, P. Tozzi and P. Rosati, Astron. Astrophys.
{\bf 398}, 879 (2003)
\bibitem{lima} J. A. S. Lima, J. V. Cunha and J. S. Alcaniz, Phys. Rev. {\bf{D68}}, 023510 (2003). astro-ph/0303388; Phys. Rev. {\bf{D69}}, 083501 (2004). astro-ph/0306319

\bibitem{ratra} G. Chen and B. Ratra, \apj {\bf{612}}, L1 (2004)

\bibitem{hogan} M. Fukugita, C. J. Hogan, P. J. E Peebles, \apj {\bf{503}}, 518 (1998)

\bibitem{simula} V. R. Eke, J. F. Navarro and C. S. Frenk, \apj {\bf{503}}, 569 (1998); J. J. Bialek, A. E. Evrard and J. J. Mohr, \apj {\bf{555}}, 597 (2001)

\bibitem{calb} R. G. Calberg {\it et al.}, Astrophys. J. {\bf{462}}, 32 (1996)

\bibitem{sne} T. Padmanabhan and T. R. Choudhury, Mon. Not. Roy. Astron. Soc.
{\bf{344}}, 823 (2003);  Z.-H. Zhu and M.-K. Fujimoto, \apj {\bf{585}}, 52 (2003); S. Nesseris and L. Perivolaropoulos, Phys. Rev. {\bf{D70}}, 043531 (2004)


\bibitem{tonry} J. L. Tonry et al., \apj {\bf{594}}, 1 (2003)

\bibitem{card} K. Freese and M. Lewis, Phys.Lett. {\bf{B540}}, 1 (2002); Z. -H. Zhu and M. -K Fujimoto, \apj, 581, 1 (2002); Z. H-. Zhu, and M. -K. Fujimoto, \apj, 602, 12 (2004); Z. H-. Zhu, M. -K. Fujimoto, and X. -T. He, \apj, 603, 365 (2004);  A. Dev, J.S. Alcaniz and D. Jain, astro-ph/0305068
                                                                                
\bibitem{clusters} S. C., Vauclair, et al., Astron. Astrop., 412, L37 (2003)






\end{thebibliography}
\end{document}